\documentclass[preprint,12pt, a4paper]{elsarticle}

\usepackage{amssymb}
\usepackage{amsmath}

\usepackage[mathlines]{lineno}

\usepackage{tikz}
\usetikzlibrary{decorations.pathreplacing}
\usetikzlibrary{shapes.geometric, arrows}

\usepackage{multirow}
\usepackage{todonotes}

\usepackage{hyperref}
\hypersetup{
    colorlinks=true,
    linkcolor=blue,
    filecolor=magenta,      
    urlcolor=cyan,
}
\usepackage{url}
\usepackage{tensor}
\usepackage{romannum}
\usepackage{xspace}
\usepackage[ruled,vlined]{algorithm2e}
\usepackage{array}
\usepackage{soul}

\newcommand{\flecsph}[1]{{\textsc{FleCSPH}\xspace#1}}
\newcommand{\flecsi}[1]{{\textsc{FleCSI}\xspace#1}}

\newcommand{\edit}[2]{\textcolor{blue}{#2}}  

\graphicspath{{}}

\journal{SoftwareX}

\begin{document}

\begin{frontmatter}



\title{\flecsph: The Next Generation FleCSIble Parallel Computational Infrastructure for Smoothed Particle Hydrodynamics}


\author{Julien Loiseau$^1$, Hyun Lim$^{1,2,3}$, Mark Alexander Kaltenborn$^{2,3,4}$, Oleg Korobkin$^{1,3}$, Christopher M. Mauney$^{3,5}$, Irina Sagert$^{2,3}$, Wesley P. Even$^{2,3,6}$, Benjamin K. Bergen$^1$}
\address{1. Applied Computer Science Group, Los Alamos National Laboratory, Los Alamos, NM 87545 USA}
\address{2. The Computational Physics and Methods Group, Los Alamos National Laboratory, Los Alamos, NM 87545 USA}
\address{3. Center for Theoretical Astrophysics, Los Alamos National Laboratory, Los Alamos, NM 87545 USA}
\address{4. Department of Physics, The George Washington University, Washington DC 20052, USA}
\address{5. HPC Environments Group, Los Alamos National Laboratory, Los Alamos, NM 87545 USA}
\address{6. Department of Physical Sciences, Southern Utah University, Cedar City, UT 84720, USA}

\begin{abstract}
\flecsph is a smoothed particle hydrodynamics simulation tool, based on the compile-time configurable framework \flecsi. 
The asynchronous distributed tree topology combined with a fast multipole method allows \flecsph to efficiently compute hydrodynamics and long range particle-particle interactions.  
\flecsph provides initial data generators, particle relaxation techniques, and standard evolution drivers, which can be easily modified and extended to user-specific setups. 
Data input/output uses the H5part format, compatible with modern visualization software. 
\end{abstract}

\begin{keyword}
Smoothed Particle Hydrodynamics \sep Tree Topology \sep High Performance Computing


\end{keyword}

\end{frontmatter}




\section{Motivation and Significance}
\label{motivation}

\flecsph is an open-source distributed smoothed particle hydrodynamics (SPH) code built on top of the Flexible Computational Science Infrastructure (\flecsi)~\cite{Bergen2017} developed at Los Alamos National Laboratory (LANL).
\flecsi is a task-based runtime abstraction layer that provides a seamless programming model for distributed-memory tasks (Legion~\cite{bauer2012legion}), and fine-grained data-parallel kernels (Kokkos~\cite{CarterEdwards20143202}), with several core topology types that can be statically specialized to support a variety of applied methods.
When using \flecsi, \flecsph has the potential to separate the application implementation from the details of machine architecture, although currently only MPI back end is implemented. 

SPH is an explicit mesh-free Lagrangian method
that solves the partial differential equations
of hydrodynamics by discretizing the flow with a 
set of fluid elements called particles~\cite{gingold77, lucy77}. 
The main SPH formula to interpolate some quantity 
$A(\vec{r})$, specified by its values over 
a set of particles $A_b \equiv A(\vec{r}_b)$, is given by
\begin{equation}
A(\vec{r}) \simeq \sum_{b\in\Omega(\vec{r})} V_b A_b W(|\vec{r}-\vec{r}_b|,h),
\end{equation}
where $W$ is a smoothing kernel, 
$h$ is the smoothing length (hydro interaction range) 
at a position $\vec{r}$, 
and $V_b$ is a volume element, usually $V_b = m_b/\rho_b$.
SPH has several advantages including handling complex geometries and support for true vacuum conditions.
Conservation of mass is included by construction, and conservation of linear momentum, angular momentum, and energy can be implemented up to machine precision. 
Since the quantities are stored in the moving particles, SPH has the advantage of exact and automatic advection.
Furthermore, the same tree structure used for determining particle neighbors can be employed for computing gravitational forces.
SPH particles can carry the stress history of the material to determine damage, model fracture, and fragmentation of solids \cite{Benz1995}.
Complicated processes in reacting flows are easily incorporated into an SPH model \cite{tartakovsky2007}.

In this paper, we outline the main features of \flecsph. The initial version of this software is described in~\cite{Loiseau2018}, while subsequent extensions are presented here.

\section{Software description}
\label{description}
\flecsph is written in C++ for UNIX computing and supercomputing platforms taking advantage of modern features of C++ and the Standard Template Library (STL)~\cite{c++2017iso}.
CMake~\cite{martin2007open} provides the build system and \flecsph can be integrated with Spack~\cite{gamblin2015spack} to construct build- and run-time environments.
Particle data is output in the H5part format \cite{adelmann2005h5part}, compatible with modern visualization software, such as SPLASH~\cite{Price2007}, Paraview~\cite{Ayachit2015} or VisIt~\cite{Childs2012}.
Simulation checkpoint and restart is supported.

\subsection{Software Architecture}
\label{architecture}
\paragraph{Drivers and initial data generators}
The main set of user ``apps'' are \emph{drivers}, which implement evolution equations, and initial data \emph{generators}, which produce initial particle configurations.
Two drivers are provided to evolve hydrodynamics with and without gravity. Users can also iterate on the set of existing drivers for more advanced physics models with different numerical evolution schemes and create custom particle system generators.
A suite of initial data generators is provided including: five standard Sod shock tubes, Sedov blast wave, Noh implosion, as well as Rayleigh-Taylor and Kelvin-Helmholtz instability.
\paragraph{Physics Functionalities}
An SPH ``particle'' is implemented as a class {\tt body\_u}, templated on the problem dimension and basic floating-point type.
`{\tt body}' refers to a single body in N-body system and `{\tt \_u}' stands for `unspecialized.'
Elements of this class consist of various particle physical properties, e.g., density and velocity, \edit{extendable to fit users' needs,}{} with standard mutators and accessors to allow customized access control.
\flecsph offers a selection of kernels, two SPH formulations, several viscosity prescriptions, particle relaxation mechanism, external conservative forces, and multiple equations of state (EOSs).
The list of options for physics functionalities
is described in the developer's \href{https://github.com/laristra/flecsph/wiki/files/notes.pdf}{notes}
on the project \href{https://github.com/laristra/flecsph/wiki}{wiki} page.
These functionalities are sorted into different C++ header files.
\paragraph{Parameter files}
The choice of physical and numerical methods is made at runtime based on the options supplied by an ASCII parameter file with a key-value syntax. 
The parameter file specifies options such as the number of particles, SPH kernels, and boundary conditions. 
The complete list of options is located on the \href{https://github.com/laristra/flecsph/wiki}{wiki} page. 
\flecsph parameter files are concise and human-readable records of simulation conditions, allowing for simple reproducibility.
\paragraph{Tree topology}
\flecsph uses a hashed tree~\cite{warren1993parallel,warren20142hot}, also known as a binary tree, quadtree, or octree in 1, 2, or 3 dimensions, respectively. 
Space-filling curves are used for the domain decomposition and the hash-table construction.
This allows for finding the parent or children of a node in $O(1)$ on average. 
Both Morton\cite{morton1966computer} (Z-order) and Hilbert space filling curves are implemented, which show faster computation of keys or better particle locality, respectively. 

\begin{figure}
\begin{center}
\begin{tabular}{cc}
    \scalebox{.95}{\begin{tikzpicture}
\begin{scope}
\def\subs{5}
\def\px{0}
\def\py{0}
\node (t1) at (0,0) [inner sep=0pt,draw,minimum width=\subs cm,minimum height=\subs cm]  {};
\pgfmathparse{\subs/2}\xdef\subs{\pgfmathresult}
\node (t10) at (t1.north west) [inner sep=0pt,anchor=north west, draw,thick,minimum width=\subs cm,minimum height=\subs cm]  {};
\node (t11) at ([xshift=-1]t10.east) [inner sep=0pt,anchor=west, draw,thick,minimum width= \subs cm,minimum height=\subs cm]  {};
\node (t12) at ([yshift=1]t10.south) [inner sep=0pt,anchor=north, draw,thick,minimum width=\subs cm,minimum height=\subs cm]  {};
\node (t13) at ([xshift=-1]t12.east) [inner sep=0pt,anchor=west, draw,thick,minimum width=\subs cm,minimum height=\subs cm]  {};
\pgfmathparse{\subs/2}\xdef\subs{\pgfmathresult}
\foreach \i in {0,2,3}
{
\node (t1\i0) at (t1\i.north west) [inner sep=0pt,anchor=north west, draw,thick,minimum width=\subs cm,minimum height=\subs cm]  {};
\node (t1\i1) at ([xshift=-1]t1\i0.east) [inner sep=0pt,anchor=west, draw,thick,minimum width= \subs cm,minimum height=\subs cm]  {};
\node (t1\i2) at ([yshift=1]t1\i0.south) [inner sep=0pt,anchor=north, draw,thick,minimum width=\subs cm,minimum height=\subs cm]  {};
\node (t1\i3) at ([xshift=-1]t1\i2.east) [inner sep=0pt,anchor=west, draw,thick,minimum width=\subs cm,minimum height=\subs cm]  {};
\draw[black!50] ([xshift=5pt,yshift=-5pt]t1\i0.north west) -- ([xshift=-5pt,yshift=-5pt]t1\i0.north east);
\draw[black!50] ([xshift=5pt,yshift=-5pt]t1\i1.north west) -- ([xshift=-5pt,yshift=-5pt]t1\i1.north east);
\draw[black!50] ([xshift=5pt,yshift=-5pt]t1\i2.north west) -- ([xshift=-5pt,yshift=-5pt]t1\i2.north east);
\draw[black!50] ([xshift=5pt,yshift=-5pt]t1\i3.north west) -- ([xshift=-5pt,yshift=-5pt]t1\i3.north east);

\draw[black!50] ([xshift=5pt,yshift=5pt]t1\i0.south west) -- ([xshift=-5pt,yshift=5pt]t1\i0.south east);
\draw[black!50] ([xshift=5pt,yshift=5pt]t1\i1.south west) -- ([xshift=-5pt,yshift=5pt]t1\i1.south east);
\draw[black!50] ([xshift=5pt,yshift=5pt]t1\i2.south west) -- ([xshift=-5pt,yshift=5pt]t1\i2.south east);
\draw[black!50] ([xshift=5pt,yshift=5pt]t1\i3.south west) -- ([xshift=-5pt,yshift=5pt]t1\i3.south east);

\draw[black!50] ([xshift=-5pt,yshift=-5pt]t1\i0.north east) -- ([xshift=5pt,yshift=5pt]t1\i0.south west);
\draw[black!50] ([xshift=-5pt,yshift=-5pt]t1\i1.north east) -- ([xshift=5pt,yshift=5pt]t1\i1.south west);
\draw[black!50] ([xshift=-5pt,yshift=-5pt]t1\i2.north east) -- ([xshift=5pt,yshift=5pt]t1\i2.south west);
\draw[black!50] ([xshift=-5pt,yshift=-5pt]t1\i3.north east) -- ([xshift=5pt,yshift=5pt]t1\i3.south west);
}
\draw[black!50] ([xshift=-10pt,yshift=-5pt]t100.north west) -- ([xshift=5pt,yshift=-5pt]t100.north west);
\draw[black!50,->] ([xshift=-5pt,yshift=5pt]t133.south east) -- ([xshift=10pt,yshift=5pt]t133.south east);

\draw[black!50] ([xshift=5pt,yshift=-5pt]t11.north west) -- ([xshift=-5pt,yshift=-5pt]t11.north east);
\draw[black!50] ([xshift=5pt,yshift=5pt]t11.south west) -- ([xshift=-5pt,yshift=5pt]t11.south east);
\draw[black!50] ([xshift=-5pt,yshift=-5pt]t11.north east) -- ([xshift=5pt,yshift=5pt]t11.south west);
\draw[black!50] ([xshift=-5pt,yshift=5pt]t103.south east) -- ([xshift=5pt,yshift=-5pt]t11.north west);
\draw[black!50] ([xshift=-5pt,yshift=5pt]t11.south east) -- ([xshift=5pt,yshift=-5pt]t120.north west);

\draw[black!50] ([xshift=-5pt,yshift=5pt]t100.south east) -- ([xshift=5pt,yshift=-5pt]t101.north west);
\draw[black!50] ([xshift=-5pt,yshift=5pt]t101.south east) -- ([xshift=5pt,yshift=-5pt]t102.north west);
\draw[black!50] ([xshift=-5pt,yshift=5pt]t102.south east) -- ([xshift=5pt,yshift=-5pt]t103.north west);
\draw[black!50] ([xshift=-5pt,yshift=5pt]t120.south east) -- ([xshift=5pt,yshift=-5pt]t121.north west);
\draw[black!50] ([xshift=-5pt,yshift=5pt]t121.south east) -- ([xshift=5pt,yshift=-5pt]t122.north west);
\draw[black!50] ([xshift=-5pt,yshift=5pt]t122.south east) -- ([xshift=5pt,yshift=-5pt]t123.north west);
\draw[black!50] ([xshift=-5pt,yshift=5pt]t123.south east) -- ([xshift=5pt,yshift=-5pt]t130.north west);
\draw[black!50] ([xshift=-5pt,yshift=5pt]t130.south east) -- ([xshift=5pt,yshift=-5pt]t131.north west);
\draw[black!50] ([xshift=-5pt,yshift=5pt]t131.south east) -- ([xshift=5pt,yshift=-5pt]t132.north west);
\draw[black!50] ([xshift=-5pt,yshift=5pt]t132.south east) -- ([xshift=5pt,yshift=-5pt]t133.north west);
\pgfmathparse{\subs/2}\xdef\subs{\pgfmathresult}
\node (t1220) at (t122.north west) [inner sep=0pt,anchor=north west, draw,thick,minimum width=\subs cm,minimum height=\subs cm]  {};
\node (t1221) at ([xshift=-1]t1220.east) [inner sep=0pt,anchor=west, draw,thick,minimum width= \subs cm,minimum height=\subs cm]  {};
\node (t1222) at ([yshift=1]t1220.south) [inner sep=0pt,anchor=north, draw,thick,minimum width=\subs cm,minimum height=\subs cm]  {};
\node (t1223) at ([xshift=-1]t1222.east) [inner sep=0pt,anchor=west, draw,thick,minimum width=\subs cm,minimum height=\subs cm]  {};
\node (t1310) at (t131.north west) [inner sep=0pt,anchor=north west, draw,thick,minimum width=\subs cm,minimum height=\subs cm]  {};
\node (t1311) at ([xshift=-1]t1310.east) [inner sep=0pt,anchor=west, draw,thick,minimum width= \subs cm,minimum height=\subs cm]  {};
\node (t1312) at ([yshift=1]t1310.south) [inner sep=0pt,anchor=north, draw,thick,minimum width=\subs cm,minimum height=\subs cm]  {};
\node (t1313) at ([xshift=-1]t1312.east) [inner sep=0pt,anchor=west, draw,thick,minimum width=\subs cm,minimum height=\subs cm]  {};
\draw (1.5,1.5) node[thick,blue] {\textbullet};
\draw ([xshift=.56 cm,yshift=-.48 cm]t101.north west) node[thick,blue,anchor=south] {\textbullet};
\draw ([xshift=.30 cm,yshift=-1.2 cm]t102.north west) node[thick,blue,anchor=south] {\textbullet};
\draw ([xshift=1. cm,yshift=-.5 cm]t121.north west) node[thick,blue,anchor=south] {\textbullet};
\draw ([xshift=.80 cm,yshift=-.7 cm]t130.north west) node[thick,blue,anchor=south] {\textbullet};
\draw ([xshift=.76 cm,yshift=-.80 cm]t120.north west) node[thick,blue,anchor=south] {\textbullet};
\draw ([xshift=.1 cm,yshift=-.1 cm]t1220.north west) node[thick,blue,anchor=north west,inner sep=0pt] {\textbullet};
\draw ([xshift=.12 cm,yshift=-.35 cm]t1223.north west) node[thick,blue,anchor=north west,inner sep=0pt] {\textbullet};
\draw ([xshift=.15 cm,yshift=-.25 cm]t1311.north west) node[thick,blue,anchor=north west,inner sep=0pt] {\textbullet};
\draw ([xshift=.15 cm,yshift=-.25 cm]t1312.north west) node[thick,blue,anchor=north west,inner sep=0pt] {\textbullet};

\end{scope}
\end{tikzpicture}} &
    \scalebox{.95}{\begin{tikzpicture}
\begin{scope}
\node[red] (1) at (-5.5,0) {1};
\node[red] (10) at (-8,-1) {10};
\node[blue] (101) at (-8.5,-2) {101};
\node[blue] (102) at (-7.5,-2) {102};
\node[blue] (11) at (-6.8,-1) {11};
\node[red] (12) at (-5.5,-1) {12};
\node[blue] (120) at (-6.5,-2) {120};
\node[blue] (121) at (-5.5,-2) {121};
\node[red] (122) at (-4.5,-2) {122};
\node[blue] (1220) at (-5,-3) {1220};
\node[blue] (1223) at (-4,-3) {1223};
\node[red] (13) at (-3,-1) {13};
\node[blue] (130) at (-3.5,-2) {130};
\node[red] (131) at (-2.5,-2) {131};
\node[blue] (1311) at (-3,-3) {1311};
\node[blue] (1312) at (-2,-3) {1312};

\draw[-] (1) -- (10);\draw[-] (1) -- (11);
\draw[-] (1) -- (12);\draw[-] (1) -- (13);

\draw[-] (10) -- (101);\draw[-] (10) -- (102);

\draw[-] (12) -- (120);\draw[-] (12) -- (121);
\draw[-] (12) -- (122);\draw[-] (122) -- (1220);
\draw[-] (122) -- (1223);

\draw[-] (13) -- (130);\draw[-] (13) -- (131);
\draw[-] (131) -- (1311);\draw[-] (131) -- (1312);
\end{scope}
\end{tikzpicture}}\\
\end{tabular}
\caption{Tree topology representation with the domain decomposition using the Z-order, the tree representation of this domain using associate keys for the entities.}
\label{fig:tree_topology}
\end{center}
\end{figure}
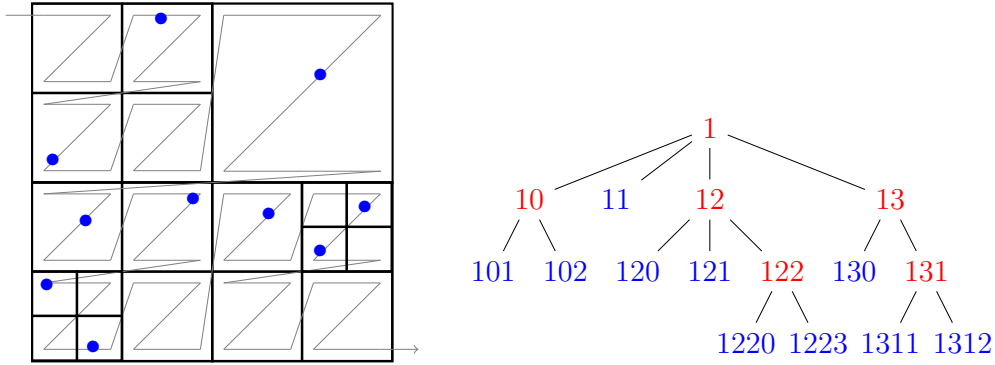

Figure \ref{fig:tree_topology} diagrams the tree topology. 
The central data structure of the tree is a hash table. 
It stores cells, denoted $hcells$, which can represent both a particle and a node. 
An $hcell$ is identified by a binary-string \emph{key} determined from the space-filling curve and the type of cell they represent: for particles, the key is computed directly from their coordinates; for nodes, it is computed based on the position of the node center of mass (CofM).
To resolve any key conflicts, each particle is assigned a unique ID.
The hashing function to distribute the $hcell$ takes the $N$ last bits of the keys, offering a perfect distribution at the bottom and contiguous storage at the top of the tree, which is accessed more often.
This representation allows for the parent or children of a branch to be easily located by adding or removing a digit in their key and accessing the hash table, an order $O(1)$ on average operation.  

Creating the distributed tree requires several steps.
First, the particles are added individually to the tree, and the necessary nodes are refined until the particles are isolated each in their own node.
The node is refined into $2^D$ sub-entities, where $D$ is the dimensionality.
After the local trees have been created, information about the top part is exchanged across the MPI ranks. 
Using a hypercube communication pattern, all the ranks share the useful data resulting in a tree where each unknown node belongs to only one other MPI rank. 

\flecsph implements hydrodynamics with SPH, and gravity with the fast multipole method (FMM). 
While SPH only computes short-range forces, gravity is a long-range interaction, requiring global communication.
Both cases of short- and long-range forces are handled by the same particle tree, but two different types of tree traversals are used: SPH and FMM.

The tree traversal is performed on a group of particles in the same node using the CofM boundary information to prune empty areas of the tree. 
A list of neighbors per particle is built, and the physics functions are called during the traversal. 
When encountering a non-local node, the ranks use asynchronous MPI to request missing information from the owner, while continuing the traversal on other particles.
When the data arrives, it is added to the tree and used to complete the neighbors list. 

\subsection{Software Functionalities}
\label{sec:soft_func}
\subsubsection{SPH Formulation in \flecsph{}}
\label{sec:sph}
\flecsph numerically solves Euler equations of ideal fluid in their Lagrangian formulation, expressing conservation of mass, momentum and energy:
\begin{eqnarray}
\frac{d \rho}{d t} &=& - \rho \nabla \cdot \vec{v}, \\
\frac{d \vec{v}}{d t} &=& - \frac{\nabla P}{\rho} + \vec{g}, \\
\frac{d u}{d t} &=& \left( \frac{P}{\rho^2} \right) \frac{d \rho}{d t},
\end{eqnarray}
where $\rho$ is density, $d/dt = \partial_t + \vec{v} \cdot \nabla$ is convective derivative, $\vec{v}$ is fluid velocity, $u$ is specific thermal energy, $P$ is pressure, and $\vec{g}$ is an external acceleration. 
The gravitational acceleration is determined by the fluid self-gravity, an external gravitational field, or both.

For SPH discretization, we use one of the simplest formulations~\cite{Rosswog2015}.
The Euler equations are discretized with the volume element $V_b = m_b / \rho_b$, and an artificial viscosity term $\Pi_{ab}$ is added:
\begin{eqnarray}
  \frac{d u_a}{dt}  
      &=& \sum_b m_b\left( 
           \frac{P_a}{\rho_a^2} + \frac12\Pi_{ab}
         \right)\vec{v}_{ab} \cdot \nabla_a W_{ab},
\label{eq:basic-dudt}
\\
  \frac{d \vec{v}_a}{d t} &=& -\sum_b m_b 
      \left( \frac{P_a}{\rho_a^2} 
           + \frac{P_b}{\rho_b^2} 
           + \Pi_{ab} \right) \nabla_a W_{ab}
       + \vec{g}_a,
\label{eq:basic-dvdt}
\end{eqnarray}
where $W_{ab} = W(|\vec{r}_a - \vec{r}_b|,h_{ab})$ and $h_{ab} = (h_a + h_b)/2$.
The density is not evolved but reconstructed from the particle positions:
\begin{eqnarray}
  \rho_a &= \sum_b m_b W_{ab}.
  \label{eq:sph-density}
\end{eqnarray}
For the viscous stress tensor $\Pi_{ab}$, we follow the standard prescription~\cite{Monaghan1992ARAA,Rosswog2015}. 
Alternatively, \flecsph{} features an implementation of the so-called thermokinetic formulation~\cite{Rosswog2015}, in which the total particle energy is evolved: $e_a = u_a + \frac{1}{2} v_a^2$. Corresponding discretized version of the energy equation reads,
\begin{eqnarray}
    \label{eqn:thermo}
    \frac{d e_a}{dt} 
      = - \sum_b m_b \left( 
             \frac{P_a \vec{v}_b}{\rho_a^2} + \frac{P_b \vec{v}_a}{\rho_b^2}
             + \frac{\vec{v}_a+\vec{v}_b}{2} \Pi_{ab}\right) \cdot \vec\nabla_a W_{ab}.
\end{eqnarray}

An adaptive smoothing length $h_a$ allows placing particles where more resolution is needed~\citep{Attwod2007AA, Cullen2010, Rosswog2015}.
The smoothing length is adapted according to the expression:
\begin{eqnarray}
\label{eqn:grad-h:adap-h}
h_a = C\eta \left( \frac{m_a}{\rho_a} \right)^{1/D}, 
\end{eqnarray}
where $D$ is dimensionality, $C$ is a kernel-dependent normalization constant, and $\eta$ is a user-specified number of neighbors~\citep{Rosswog2015}.
In this method, the number of neighbors for hydrodynamic interactions remains approximately constant for all particles and times.

\subsubsection{Computing Gravitational Force via FMM}
\label{sec:fmm}

\begin{figure}[ht!]
  \centering
  \begin{tabular}{cc}
  \includegraphics[width=0.52\textwidth]{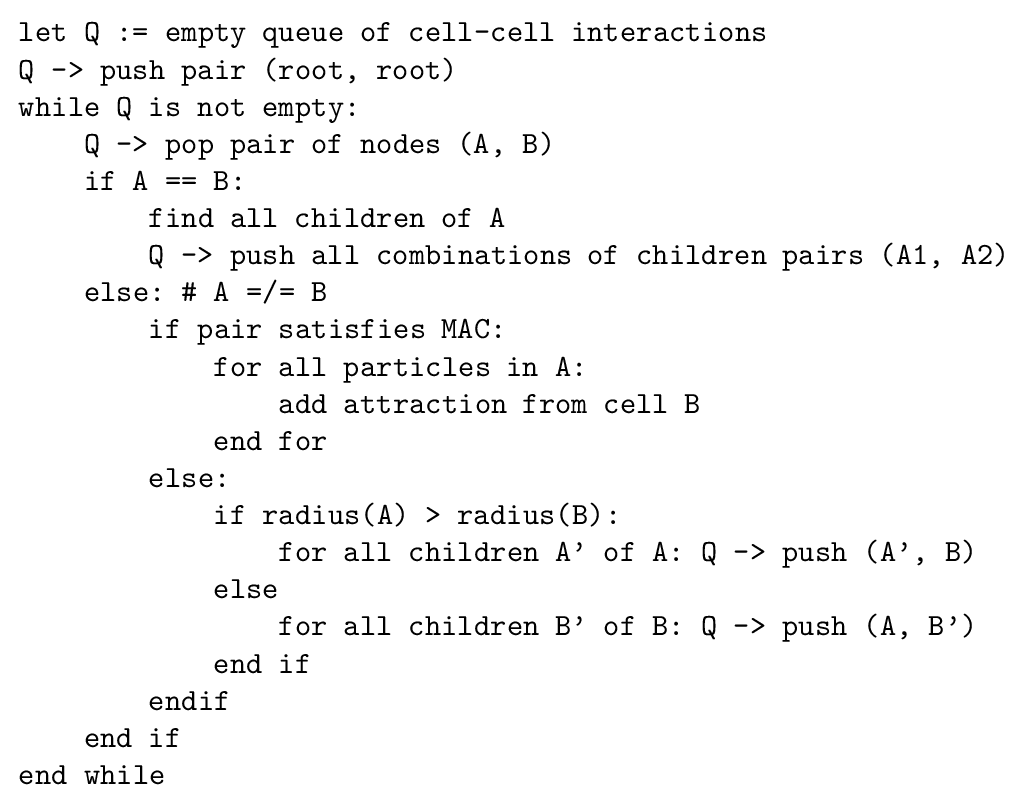} &
  \includegraphics[width=0.44\textwidth]{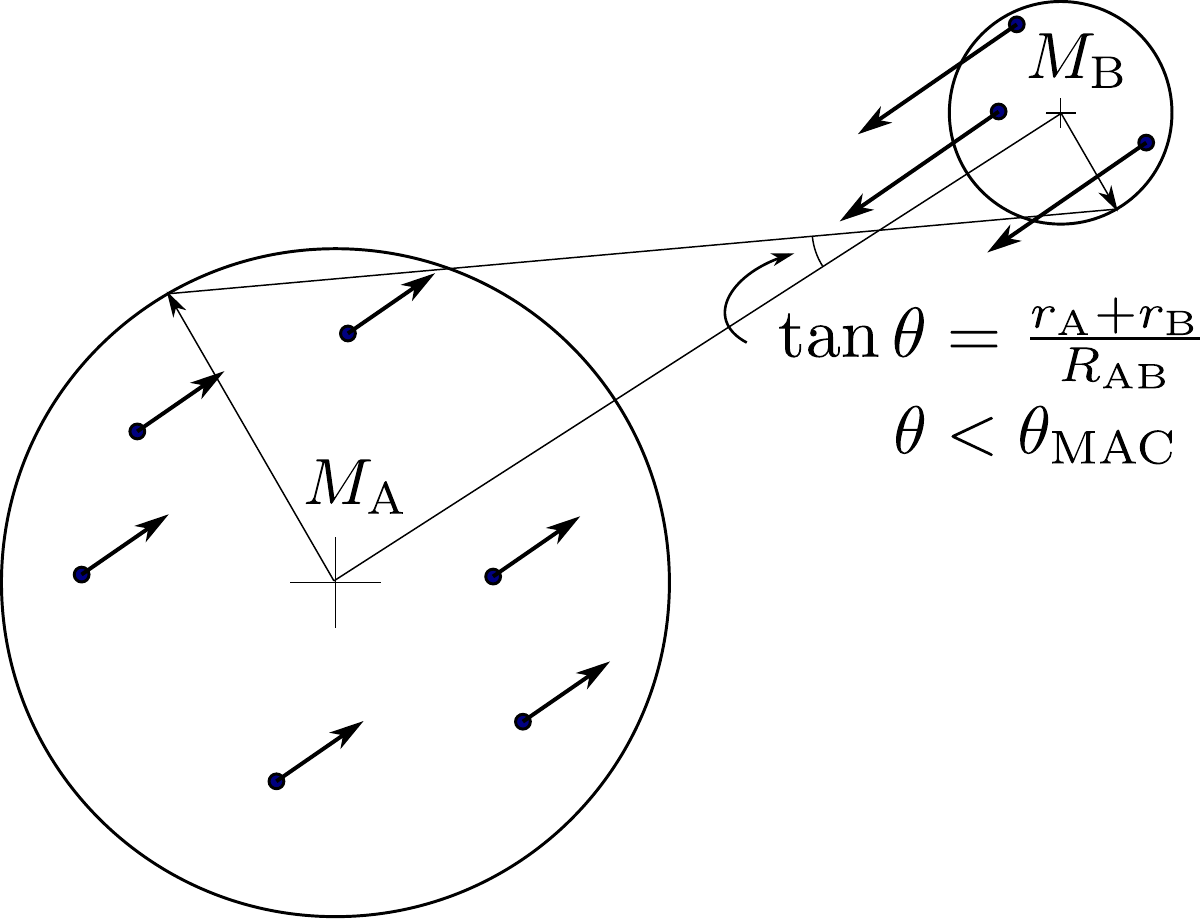}
  \end{tabular}
  \caption{Fast Multipole Method: pseudocode algorithm and an illustration of the
  symmetric gravitational interaction between two nodes.
  Node proximity is quantified by angle $\theta$: if $\theta > \theta_{\rm MAC}$, larger
  node is split up into child nodes to refine the interaction.
  } 
  \label{fig:fmm}
\end{figure}

\flecsph{} uses the FMM approximation~\cite{Greengard1987} to treat gravitational interactions, following~\citep{Dehnen2000ApJ}, up to the first order in the Taylor expansion. 
Without approximation, pairwise N-body interactions have $O(N^2)$ computational complexity. 
FMM replaces gravitational forces between individual particles in distant nodes with symmetrized node-node interactions~\cite{Dehnen2000ApJ}.
Nodes are accepted for node-node interaction, if they satisfy the so-called ``Minimal Acceptance Criterion'' (MAC, see Figure~\ref{fig:fmm}).

Attraction of a distant node is approximated by a series of multipoles.
Because of the symmetry in nodal interactions, FMM conserves linear momentum.
In the implemented first order of the multipole expansion, angular momentum is conserved exactly.

The gravitational acceleration $\vec{g}_a$ of a particle $a$ is computed as the sum of attractions from all cells passed during the tree traversal while satisfying the MAC angle with respect to the particle:
\begin{equation}
\label{eqn:grav}
\vec{g}_a = -\sum_{{}_{{\rm MAC}(A, B)}} \frac{G M_B}{|\vec{R}_{AB}|^3}\vec{R}_{AB},
\end{equation}
where $G$ is the Newton's gravitational constant, $A$ is the cell containing particle $a$, $a\in A$, and $B$ is the other cell, $a\not\in B$.
$\vec{R}_A$ and $\vec{R}_B$ are the centers of mass for $A$ and $B$, respectively. $\vec{R}_{AB} = \vec{R}_A - \vec{R}_B$ is the distance vector, and $M_B$ is the mass of $B$.

When $\theta_{\rm MAC} = 0$, eq.~(\ref{eqn:grav}) reduces to the exact expression for Newtonian interactions between individual particles. 

\section{Illustrative Examples}
\label{sec:examples}
%

In this section, we provide several test cases to demonstrate 
validation and functionalities. Detailed instructions for all test cases 
can be found in the 
\href{https://github.com/laristra/flecsph/wiki}{\texttt{wiki}} page.

\subsection{Basic Hydrodynamics Problems: Sod Shock Tube Test}
The Sod shock tube is a standard test with a classical Riemann problem with the following initial parameters:
\begin{equation}
(\rho, v, p)_{t=0} = 
\begin{cases}
(1.0,0.0,1.0) & \text{if} \indent 0.0 < x \leq 0.5 \\
(0.125,0.0,0.1) & \text{if} \indent 0.5 < x < 1.0.
\end{cases}
\end{equation}
This leads to the development of a shock front, which propagates from high-density into low-density regions, and followed by a contact discontinuity. A density rarefaction wave propagates into the high-density region. 

\begin{figure}[ht!]
\begin{center}
\includegraphics[width=.8\linewidth]{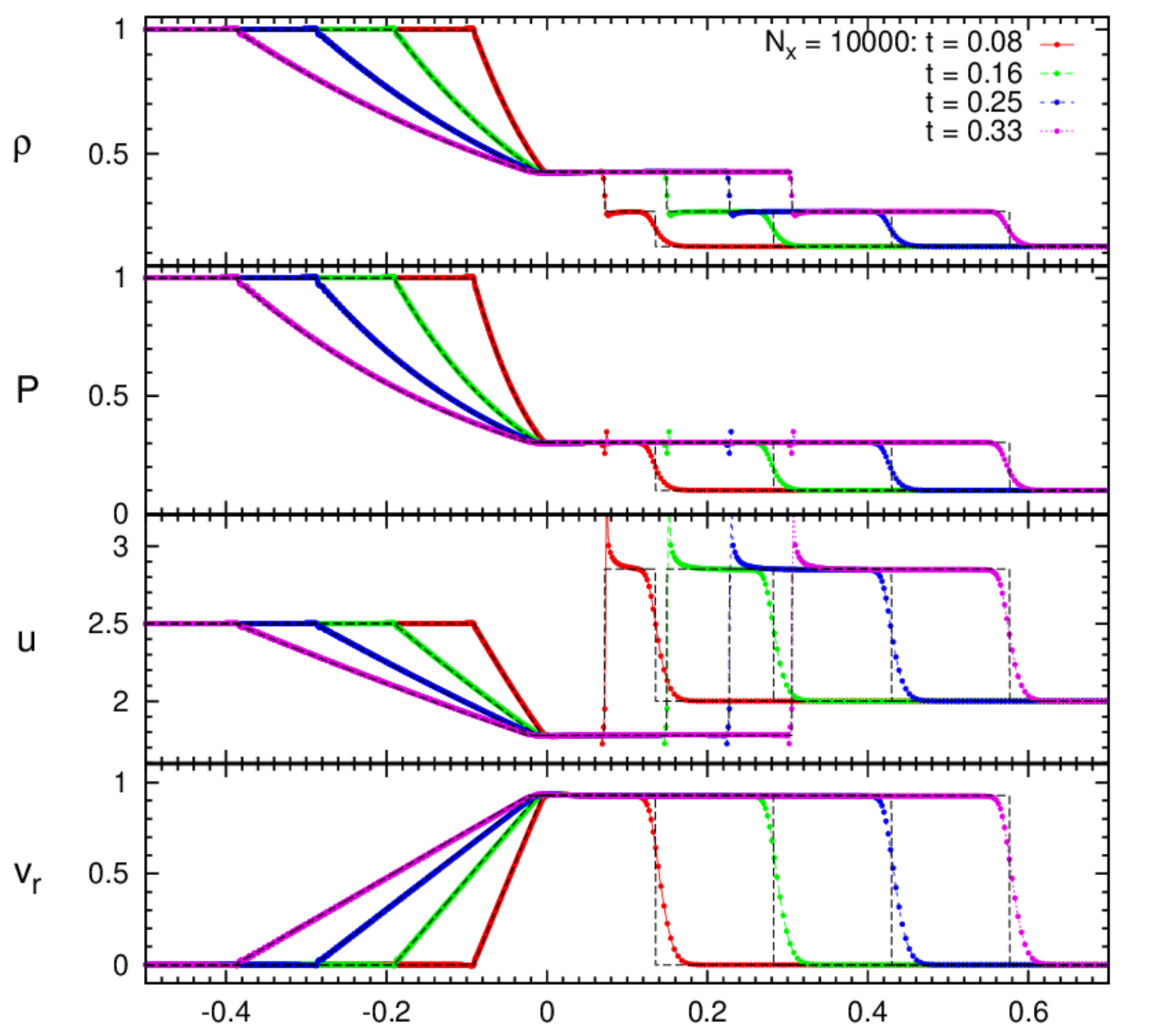}
\caption{One-dimensional Sod shock tube with 10,000 particles. 
Panels, from top to bottom: density, pressure, specific internal energy, and velocity.
Each panel contains four different times. 
Dashed black lines show analytic solution.}
\label{fig:sod}
\end{center}
\end{figure}

Figure~\ref{fig:sod} shows a 1D Sod shock tube with
$10,000$ particles. 
Density($\rho$), pressure($P$), specific internal energy($U$), and velocity($v_r$) are plotted 
at four different times.
The dashed black lines are analytic solutions.
Despite small deviations appearing near shock and contact discontinuities, our results agree well with the analytic solution.

\subsection{Testing Gravity: Stellar Oscillations}
\label{sec:fmm_test}

\begin{figure}[ht!]
  \begin{tabular}{cc}
  \includegraphics[width=0.5\textwidth]{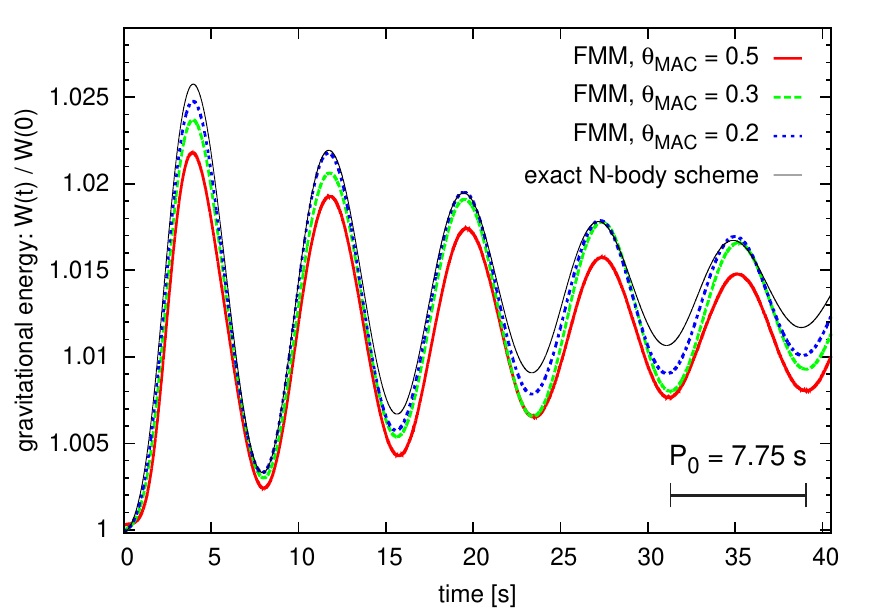} &
  \includegraphics[width=0.5\textwidth]{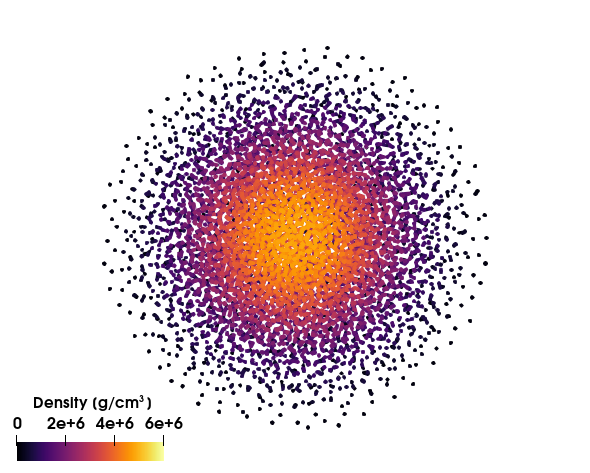}
  \\
  \includegraphics[width=0.5\textwidth]{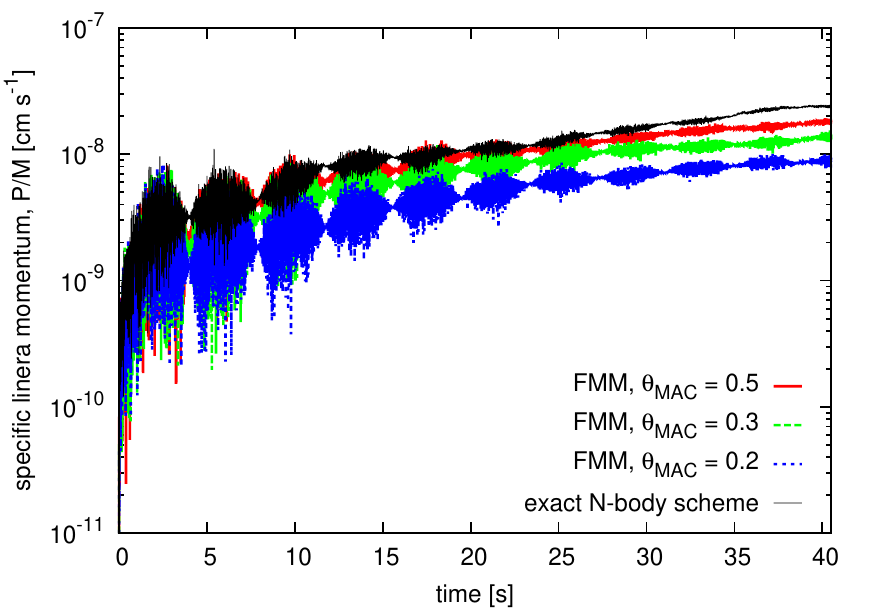} &
  \includegraphics[width=0.5\textwidth]{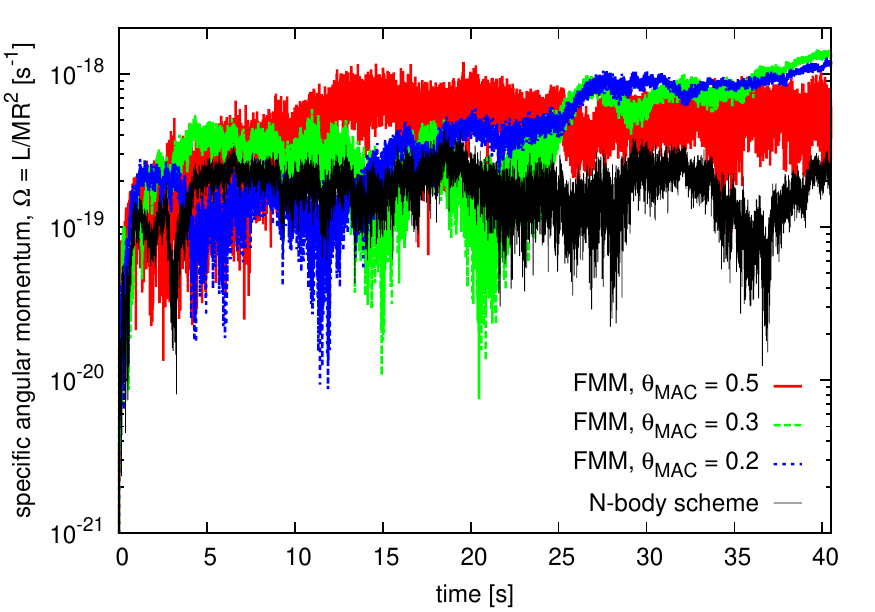}
  \end{tabular}
  \caption{Oscillations of a star near equilibrium (14,993 particles).
  Top left panel: gravitational energy evolution for the exact N-body gravity and the FMM approximation with three different MAC values: $\tan{\theta_{\rm MAC}} = 0.2$, $0.3$ and $0.5$.
  Top right panel: 3D rendering of particle positions with ParaView.
  Bottom row: evolution of specific linear (right) and angular momenta (left).
  } 
  \label{fig:fmm_single_wd}
\end{figure}

As a demonstration of the numerical methods for self-gravitating fluids, we present evolution of a stable isolated star in equilibrium.
Truncation error in the initial configuration triggers small oscillations of the star, most notably its fundamental radial mode and the first few overtones \cite{cox80}.
These oscillations are damped by the viscosity during the evolution.
This test checks consistency and conservation properties for the coupled hydrodynamics and gravity.
We compare conservation of energy, momentum, and angular momentum, computed using the FMM approximation, and using the exact pairwise N-body particle interactions with $O(N^2)$ complexity.
For the initial data, we solve Lane-Emden equation~\cite{chandra1957} 
for polytrope with $\Gamma=5/3$, $K = 10^{12}$ (in CGS units), and central density $\rho_c =5.2\times10^{6}\,{\rm g}\,{\rm cm}^{-3}$. 
This results in a polytrope resembling a white dwarf, with mass $0.2\,M_\odot$ and radius 4790~km.
and period of adiabatic oscillations in the fundamental mode for this kind of polytrope is $7.75$~s (see e.g. \cite{cox80}, Table 8.1).

Figure~\ref{fig:fmm_single_wd} shows a star discretized with 14,993 equal-mass particles evolved using the thermokinetic formulation.
It demonstrates perfect conservation of linear and angular momenta to machine precision.
The top right panel shows the relaxed particle configuration.
The top left panel displays the evolution of the gravitational energy (normalized to its initial value), for the exact N-body gravity with pairwise interactions, 
and FMM implementation with different values of $\tan{\theta_{\rm MAC}}$.
Gravitational energy oscillates with the fundamental frequency.
Evolution of the gravitational energy differs for the different values of MAC, but the results approach the N-body case when the MAC angle decreases.
Since the $O(N^2)$ algorithm computes gravitational forces exactly, this provides validation of the FMM implementation. 
The FMM algorithm could be extended to higher orders in Taylor expansion~\cite{Dehnen2000ApJ,Dehnen2014} to provide better agreement with the exact N-body scheme.
However, conservation of angular momentum breaks down for higher orders, and special techniques are needed to recover it \cite{Marcello2017}.

\subsection{Integrate Example: White Dwarf Binary}
\label{sec:binary}

\begin{figure}[ht!]
  \centerline{\includegraphics[width=0.9\textwidth]{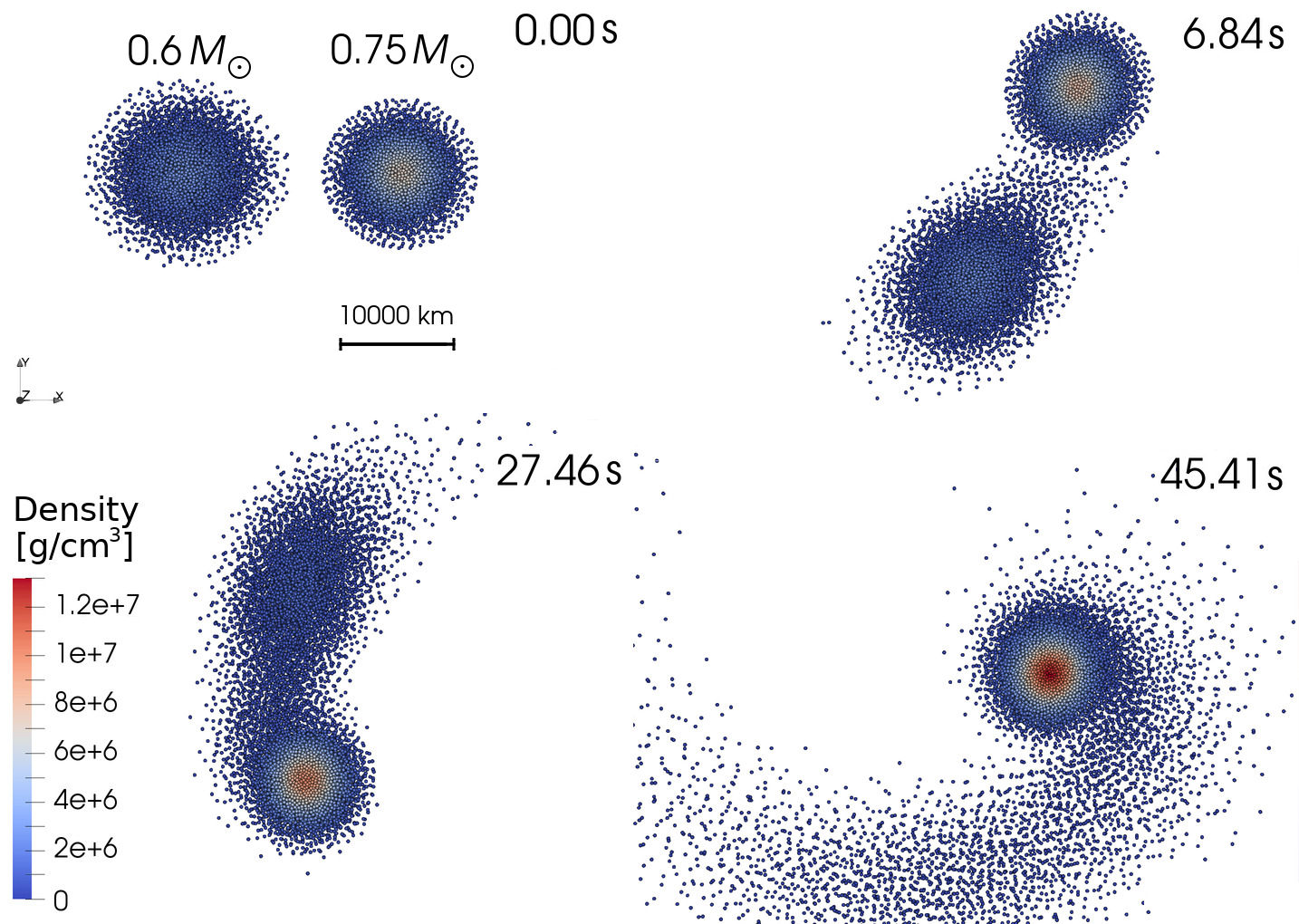}}
  \caption{The density of particles in a binary white dwarf merger. 
  } 
  \label{fig:bwdmerger}
\end{figure}
A prominent application for SPH is 
stellar mergers~\cite{Rasio1995, Dan2011, Rosswog2013, Motl2017}. Here, we present a binary white dwarf(WD) simulation. To set up the system spherical particle distributions are generated for the individual WDs, similar to the setup in Section~\ref{sec:fmm_test}.
The configurations are then placed on a Keplerian orbit.  

Figure~\ref{fig:bwdmerger} shows four frames of the merging double
WD binary. The co-rotating system is composed of 21,295 particles using the zero temperature WD EOS~\citep{chandra1935} and has an initial period of 40.1$s$. 
In order to limit the simulation time for this example the particle number has been lowered and the orbit is chosen such that the stars immediately being transferring mass and merge within a single orbital period.  In nature the stars would begin to slowly transfer mass at a wider orbit and this process would evolve over many orbital periods, but end a qualitatively similar state. 

\subsection{Scalability}
\label{sec:scalability}

\begin{figure}[h!]
\begin{tabular}{cc}
\includegraphics[width=.5\textwidth]{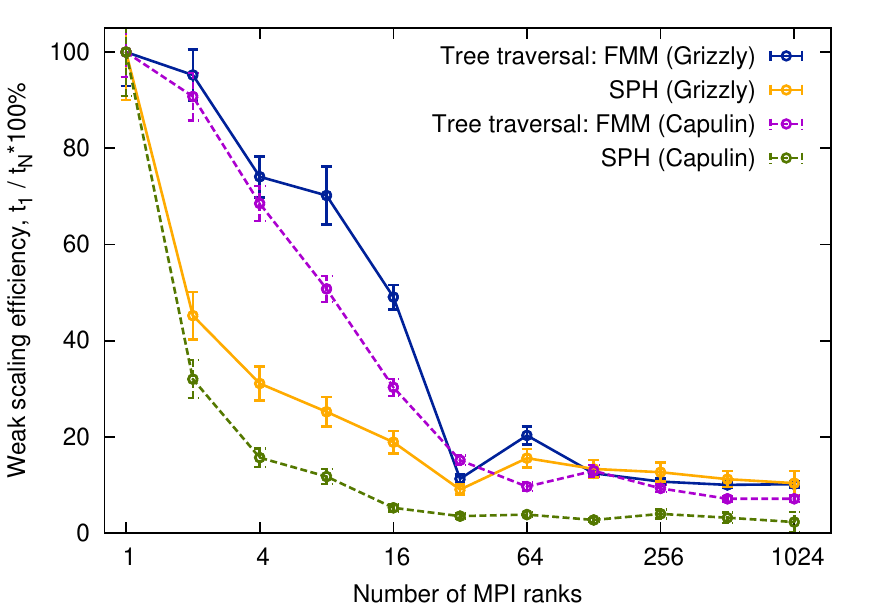} &
\includegraphics[width=.5\textwidth]{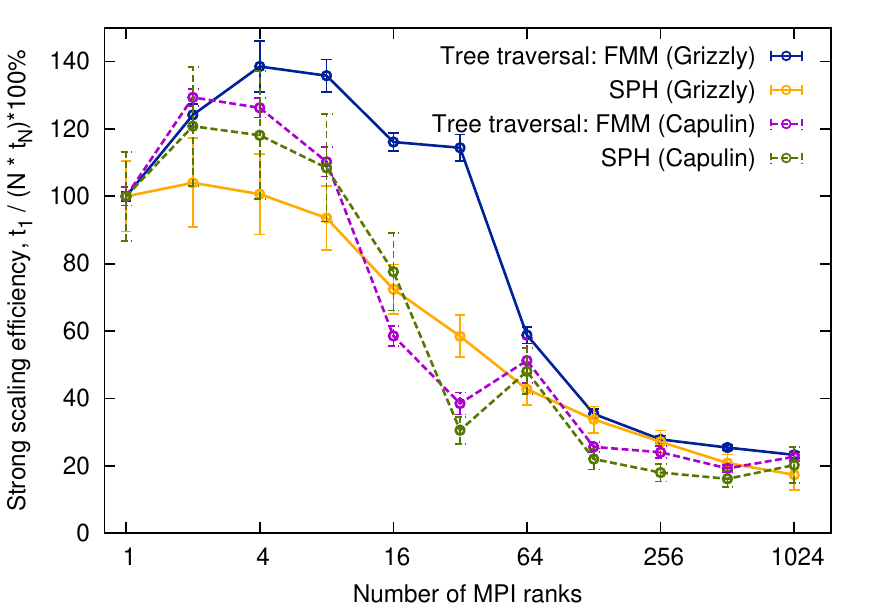}
\\
\includegraphics[width=.5\textwidth]{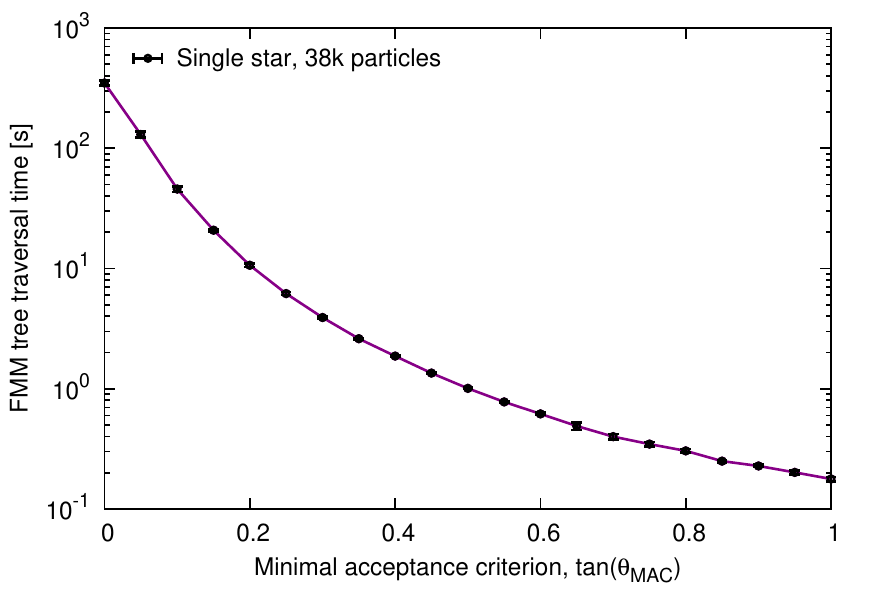} &
\includegraphics[width=.5\textwidth]{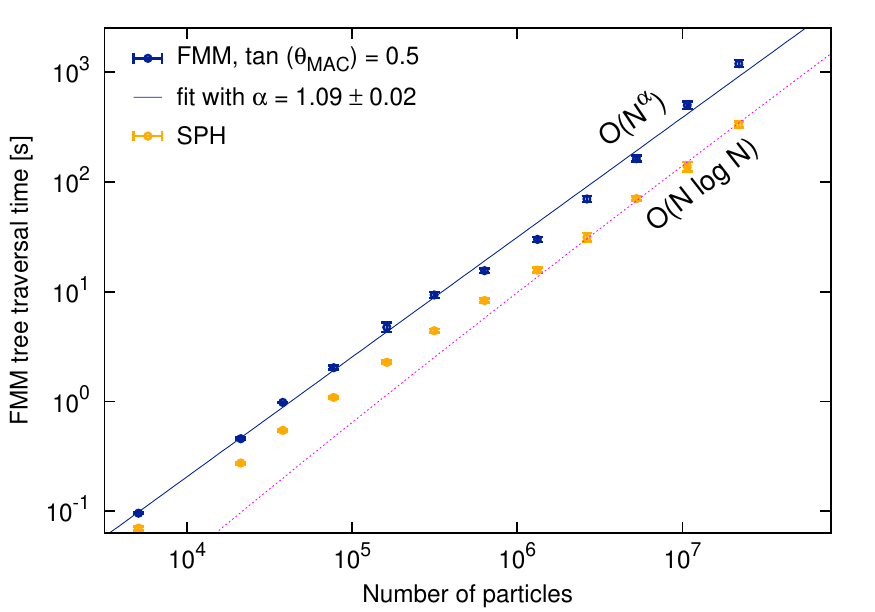}
\end{tabular}
\caption{Top row: weak (left) and strong (right) scaling efficiency for FMM and SPH tree traversal time, on two clusters (Grizzly and Capulin). 
Weak scaling used $5000$ particles per rank and strong scaling used $10^7$ particles.
Bottom left: time spent per iteration as a function of MAC.
Bottom right: timing for a single FMM and SPH tree traversal on a single node as a function of the number of particles $N$, as well as fits with power law $N^\alpha$ and $N\log{N}$. 
Fits demonstrate asymptotic complexity of $O(N)$ for FMM and $O(N\log{N})$ for SPH tree traversal implementations.
} 
\label{fig:scaling}
\end{figure}

Figure~\ref{fig:scaling} demonstrates strong and weak scaling of \flecsph{} on LANL supercomputing clusters Grizzly and Capulin. 
Grizzly is an 8-SU cluster running RHEL Linux v.7.7, it has dual socket 2.1 GHz 18 core Intel Broadwell E5 2695v4 processor with 45MB of cache and 128GB of RAM on each node.
Capulin is an Advanced RISC Machine cluster (ARM) by Cray, Inc., with 56 cores in simultaneous multi-threading (SMT) mode with four threads per core and 256GB of RAM per node. Scaling runs on Capulin were capped at 32 cores/node.

The oscillating star setup with different numbers of particles was used for both tests (see Section~\ref{sec:fmm_test}). 
For jobs with number of ranks from 1 to 32, we used a single node, and for larger jobs -- multiple nodes with 32 ranks per node.
This manifested in a drop in strong scaling efficiency on Grizzly between 32 and 64 ranks is due to the inter-node data transfer. On Capulin, efficiency degrades due to increased context switching between SMT threads.
For the weak scaling, efficiency drops between 1 and 2 ranks due to parallel overhead, but then remains relatively flat.
Overall both SPH and FMM tree traversals show comparable scaling.

Unlike the exact N-body with quadratic complexity, FMM algorithm has linear (or even sublinear) complexity\citep{Dehnen2014}.
For \flecsph{}, this is shown in the bottom right panel of Fig.~\ref{fig:scaling}.
The SPH tree traversal is expected to have $O(N\log{N})$ complexity, which is satisfied in \flecsph{} for the highest number of particles.

The timing of FMM algorithm strongly depends on the MAC angle.
Top left panel in Fig.~\ref{fig:scaling} shows the timing of a single FMM tree traversal as a function of MAC.
In the extremes, MAC=0 reduces to the exact N-body, while MAC=1 corresponds to nodes in contact, i.e., accepting any nodes that do not intersect.
The speedup between the extremes for 38,000 particles is almost three orders of magnitude, which will be even higher for larger number of particles.
This comes at a moderate price, because using higher MAC reduces code accuracy, as shown in Fig.~\ref{fig:fmm_single_wd}. 

\section{Impact \& Conclusion}
\label{sec:impact}

\flecsph is designed to be a performance portable particle hydrodynamics simulation tool, oriented to explore modern heterogeneous architectures and massive parallelism.
It opens an easy avenue for researchers to write efficient applications that will perform at scale, for those areas that can benefit from meshfree methods.
Its modular design allows users to extend the initial hydrodynamics + gravity suite with a variety of other multi-physics applications.
In this work, we demonstrate the structure and capabilities of \flecsph using several examples of standard hydrodynamic tests and a few astrophysical applications. 

FleCSPH was designed from conception to use the compile-time configurable framework FleCSI, with its functional programming model for execution, control, and data abstractions that are consistent both with MPI and with task-based runtime systems such as Legion (distributed) or Kokkos (node-level).
However, the current version of FleCSPH does not take advantage of all features provided in FleCSI: it is limited to only having the support for the MPI backend.
Future development plans include incorporating more FleCSI functionalities into FleCSPH.

As its nearest goal, \flecsph will address problems in astrophysics that involve highly irregular morphologies and are sensitive to conservation properties.
Examples of such problems include mergers of binary white dwarfs or neutron stars, tidal disruptions of stars by black holes, fallback accretion, planetary impacts, and more.
FleCSPH is currently being used by scientist at LANL to address these research topics and is also being used as an educational tool for participants in LANL's computational science and physics student programs.

With the recent advances in SPH techniques, new codes have been developed \cite{Price2018, Owen2001, Rosswog2019b}, bringing the fidelity needed to resolve some of these difficult research problems.
\flecsph will feature performance portability in an open-source environment, which will allow its users to study the big open questions in astrophysics and hydrodynamics at scale.
Several natural avenues for future development include new multi-physics applications, such as Lagrangian magnetohydrodynamics~\cite{Price2012}, radiation transport in the flux-limited diffusion approximation~\cite{Whitehouse2004}, and coupling to other methods such as Monte Carlo method~\cite{Oxley2003}. 

\section*{Acknowledgements}
\label{acknowledgements}
The authors thank Gary Dilts, Chris Fryer, Pascal Grosset, Christoph Junghans, Jonah Miller, Nick Moss, Brett Okhyusen and Stephan Rosswog for discussions. 
This work was supported by the LANL ASC Program and LDRD grants 20190021DR and 20200145ER. The development used LANL Institutional Computing Program resources. LANL is operated by Triad National Security, LLC, for the National Nuclear Security Administration of the U.S.DOE (Contract No. 89233218CNA000001). 



\bibliographystyle{elsarticle-num} 
\bibliography{refs}

\newpage
\section*{Current code version}
\label{sec:code_version}

\begin{table}[!h]
\begin{tabular}{|l|p{6.5cm}|p{6.5cm}|}
\hline
\textbf{Nr.} & \textbf{Code metadata description} & \textbf{Value} \\
\hline
C1 & Current code version & Version 1.2 \\
\hline
C2 & Permanent link to code/repository used for this code version & \url{github.com/laristra/flecsph} \\
\hline
C3 & Legal Code License   & BSD 3-Clause License \\
\hline
C4 & Code versioning system used & git \\
\hline
C5 & Software code languages, tools, and services used & C++17, MPI, OpenMP. \\
\hline
C6 & Compilation requirements, operating environments \& dependencies & OS: Linux/OSX. DEP: FleCSI, HDF5, GSL \\
\hline
C7 & Link to developer documentation/manual & \url{github.com/laristra/flecsph/wiki} \\
\hline
C8 & Support email for questions & flecsph-support@lanl.gov\\
\hline
\end{tabular}
\caption{Code metadata}
\label{table} 
\end{table}

\end{document}